\documentclass[sigconf]{acmart}
\AtBeginDocument{%
  \providecommand\BibTeX{{%
    \normalfont B\kern-0.5em{\scshape i\kern-0.25em b}\kern-0.8em\TeX}}}

\copyrightyear{2023}
\acmYear{2023}
\setcopyright{acmlicensed}\acmConference[CHCHI 2023]{Chinese CHI
2023}{November 13--16, 2023}{Denpasar, Bali, Indonesia}
\acmBooktitle{Chinese CHI 2023 (CHCHI 2023), November 13--16, 2023,
Denpasar, Bali, Indonesia}
\acmPrice{15.00}
\acmDOI{10.1145/3629606.3629643}
\acmISBN{979-8-4007-1645-4/23/11}

\begin{document}

\title{VR PreM+: An Immersive Pre-learning Branching Visualization System for Museum Tours}

\author{Ze Gao}
\authornote{Both authors contributed equally to this research.}
\email{zgaoap@connect.ust.hk}
\affiliation{%
  \institution{Hong Kong University of Science and Technology}
  \city{Hong Kong SAR}
  \country{China}
}

\author{Xiang Li}
\authornotemark[1]
\email{xl529@cam.ac.uk}
\affiliation{%
  \institution{University of Cambridge}
  \city{Cambridge}
  \country{United Kingdom}
}

\author{Changkun Liu}
\email{cliudg@connect.ust.hk}
\affiliation{%
  \institution{Hong Kong University of Science and Technology}
  \city{Hong Kong SAR}
  \country{China}
}

\author{Xian Wang}
\email{xian.wang@connect.ust.hk}
\affiliation{%
  \institution{Hong Kong University of Science and Technology}
  \city{Hong Kong SAR}
  \country{China}
}

\author{Anqi Wang}
\email{awangan@connect.ust.hk}
\affiliation{%
  \institution{Hong Kong University of Science and Technology}
  \city{Hong Kong SAR}
  \country{China}
}

\author{Liang Yang}
\email{lyangbl@connect.ust.hk}
\affiliation{%
  \institution{Hong Kong University of Science and Technology}
  \city{Hong Kong SAR}
  \country{China}
}

\author{Yuyang Wang}
\email{yuyangwang@hkust-gz.edu.cn}
\affiliation{%
  \institution{Hong Kong University of Science and Technology (Guangzhou)}
  \city{Guangzhou}
  \country{China}
}

\author{Pan Hui}
\email{panhui@ust.hk}
\affiliation{%
  \institution{Hong Kong University of Science and Technology (Guangzhou)}
  \city{Guangzhou}
  \country{China}
}

\author{Tristan Braud}
\email{braudt@ust.hk}
\affiliation{%
  \institution{Hong Kong University of Science and Technology}
  \city{Hong Kong SAR}
  \country{China}
}

\renewcommand{\shortauthors}{Gao and Li, et al.}

\begin{abstract}

We present VR PreM+, an innovative VR system designed to enhance web exploration beyond traditional computer screens. Unlike static 2D displays, VR PreM+ leverages 3D environments to create an immersive pre-learning experience. Using keyword-based information retrieval allows users to manage and connect various content sources in a dynamic 3D space, improving communication and data comparison. We conducted preliminary and user studies that demonstrated efficient information retrieval, increased user engagement, and a greater sense of presence. These findings yielded three design guidelines for future VR information systems: display, interaction, and user-centric design. VR PreM+ bridges the gap between traditional web browsing and immersive VR, offering an interactive and comprehensive approach to information acquisition. It holds promise for research, education, and beyond.
\end{abstract}





\maketitle

\section{Introduction}

\begin{figure*}[ht]
    \centering
    \includegraphics[width=\linewidth]{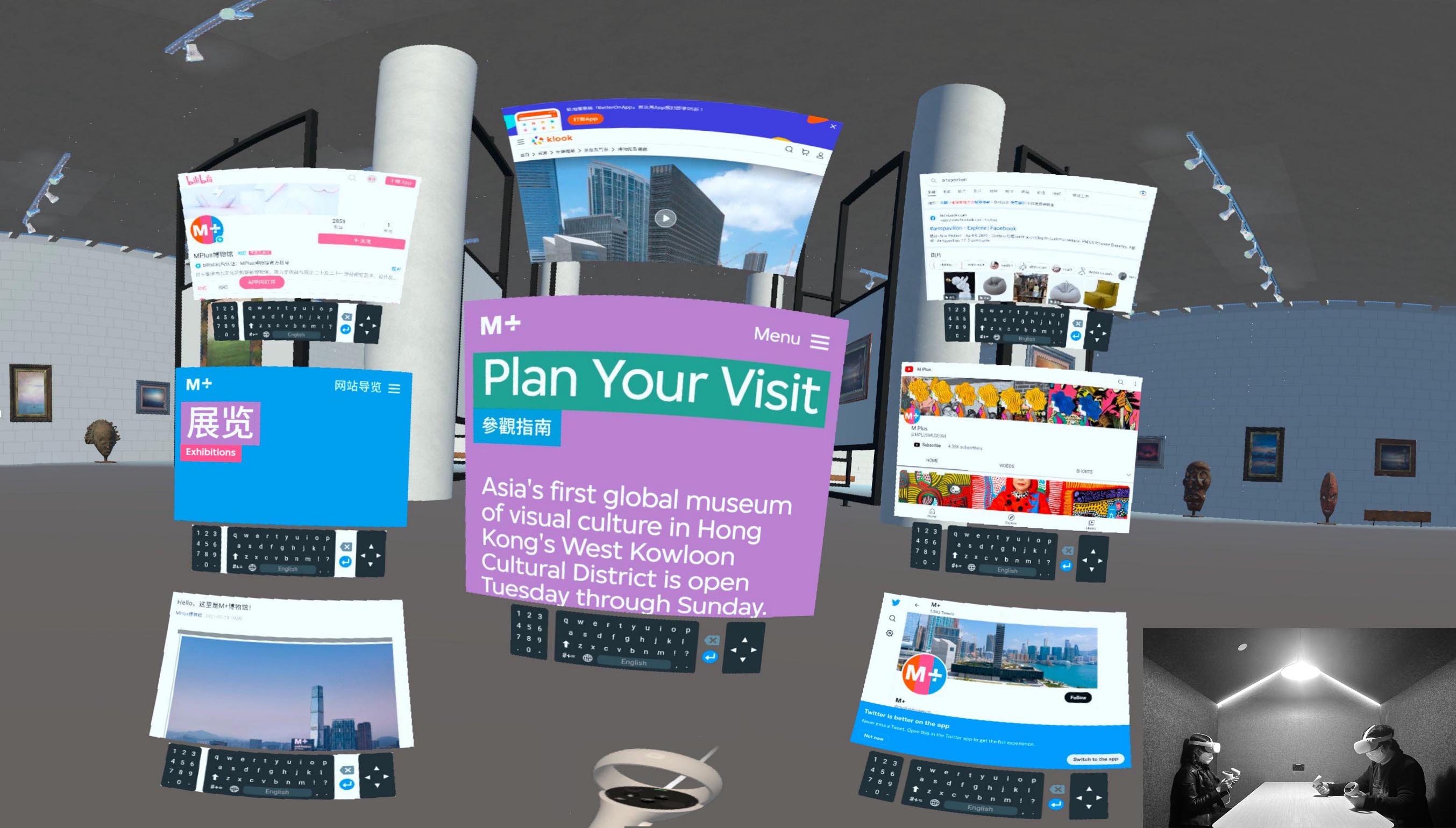}
    \caption{We present the VR PreM+, a branching system for information acquisition through interactive virtual reality environments.}
    \Description{Two users using the VR PreM+ Branching System for information acquisition.}
\label{fig:teaser}
\end{figure*}

The COVID-19 pandemic has been forcing us to perform many daily activities online. Immersive technologies such as augmented reality (AR) and virtual reality (VR) have shown encouraging capabilities for performing such activities. However, their pervasive nature requires us to rethink information gathering~\cite{lam2019m2a} and learning processes~\cite{pokhrel2021literature} to maximize their utility. While much research is dedicated to education in VR, few works on collaborative and interactive systems facilitate the preparatory activities before formal learning, also called pre-learning. Whether exploring new concepts or enhancing the understanding of an already-known topic, students and instructors require significant access to information.
Currently, information is limited by the access medium.
For example, desktop browsers arrange web pages as tab lists that complicate sorting and classifying information.
While there are information organization tools like Miro\footnote{\url{https://miro.com}} to create associations and hierarchies, they are still limited to the two-dimensional (2D) space of the computer screen\cite{gao2023envisioning,wongthird}.
Immersive technologies help us present information conveniently by leveraging the third dimension to organize chunks of information~\cite{roberts20213D,balakrishnan2006aren}. However, current web browsers in AR and VR are heavily inspired by 2D web browsers and do not fully leverage the potential of the 3D space for information classification and presentation. 


In this paper, we fully develop VR PreM+, an immersive VR interface to browse, extract, and classify information on the Web as an extension and build-up of previous envisioning\cite{gao2023envisioning,wang2022envisioning} and the formative study of an Immersive Multi-Screen VR System\cite{wongthird}. 

VR PreM+ is based on the observation that digital users often arrange information and applications side by side to improve productivity.
However, whether splitting the screen on mobile devices or multiplying the screens on desktop computers, the available display area is limited by physical space.
Our VR pre-learning system avoids these issues by leveraging the near-unlimited virtual space in VR. VR PreM+ places the users in a learning room that guides the pre-learning process.

In this room, users can spawn multiple windows and arrange them in the 3D space to better draw relationships between pieces of information. As such, VR PreM+ addresses the need for personalization and browsing classification for web-based information 
retrieval in pre-learning. 
We design VR PreM+ with the following concepts in mind:

\begin{enumerate}
\setlength\itemsep{-0.1em}
    \item \textbf{Functionality:} The field of information retrieval in VR environments is still in its infancy, and it is still unclear what interfaces facilitate this process. A few studies focus on 3-dimensional information retrieval in virtual spaces, while most existing work on information retrieval remains on 2-dimensional screens.
    \item \textbf{Validity:} Personalized draggable interfaces in 3D environments still lack systematic experimental evaluation. We explore whether information retrieval in a virtual environment without spatial constraints has higher efficiency than the 2D screen interface.
    \item \textbf{Humanized design:} Besides user performance, we also consider subjective metrics such as perceived quality and perceived task load, and qualitative interviews to promote humanistic VR environments.
    \item \textbf{Collaboration \& Interaction:} The information collected through VR PreM + can be viewed on any end device, allowing users to collaborate between devices. Users can share information about the same topic even if they are not using the same device.
\end{enumerate}

Noting the success of VR for museum exhibitions (over 2000 museums in Google Arts \& Culture), we consider the lack of tools for users to gather information before their visit. Preparing for museum visits can be seen as a pre-learning task where users browse diverse digital media (text, photos, videos, 3D objects, maps) from different sources to explore what the museum offers and plan practical aspects.
Thus, we apply VR PreM+ to a museum visit scenario. Users gather information on the museum and its exhibitions in a VR environment and sort it according to their personal preferences, regardless of the type of media.


To better understand the limitations of current 2D and 3D infor-mation-gathering methods, we first conduct a formative study (N=24) where we ask participants to perform the pre-learning process using a traditional desktop web browser and Oculus default VR web browser. Although the apparatus does not significantly influence the user experience quality and the perceived workload, users were significantly faster when using traditional 3D browsers. Participants also reported that the default VR browser was less convenient to use than its 2D counterpart, from which it draws inspiration. Given these results, we design VR PreM+ to facilitate adequate access to information. We evaluate VR PreM+ over a panel of 18 participants and compare it to the 2D and VR default web browsers. VR PreM+ shows significantly higher hedonic quality and presence than the other methods. Participants appreciated the system's personalization capabilities and the larger display space enabled by the VR environment to sort information. Our contribution can be summarized as follows: 

\begin{enumerate}
\setlength\itemsep{-0.1em}
\item \textbf{Formative study on VR information browsing techniques.} Users prefer desktop 2D browsers to current 3D VR browsers.
\item \textbf{Design and Implementation of VR PreM+}, a VR pre-learning branching visualization system for museum tour planning.
\item \textbf{Quantitative and qualitative evaluation} of VR PreM+ through a user study (N=18). VR PreM+ halves ($348\,s$) the time to retrieve information and significantly increases the hedonic quality and presence while maintaining the same task load compared to traditional and VR web browsers.
\item \textbf{Formulation of strategies} to guide the design of information retrieval systems in VR.
\item \textbf{Flexible multi-channel approach.} Allowing users to focus on information from one channel when needed selectively, or the flexibility to adjust multiple channel feeds to the most convenient area for comparative viewing.
\end{enumerate}

The rest of this paper is organized as follows. After summarizing the existing works in the field of information retrieval systems in Section~\ref{sec:related}, we develop the application scenario in Section~\ref{sec:scenario}. We then proceed to the formative study in Section~\ref{sec:formative}, which results in guide the design of VR PreM+ in Section~\ref{sec:design}. We finally conduct a user evaluation in Section~\ref{sec:evaluation} and develop design strategies and limitations in Section~\ref{sec:strategies} and Section~\ref{sec:limitations}.

\section{Related Work}
\label{sec:related}

Our research focuses on applying VR for information retrieval and comparative analysis of information prioritization in the era of the metaverse to determine the most appropriate tour path in the shortest time. Related research areas in museums include existing information retrieval systems, VR systems with information retrieval, and the current state of visualization branching systems in VR.

\subsection{Information Retrieval Systems}
Information retrieval (IR) is the process of searching for relevant information~\cite{campos2014survey}, which refers to the search for unstructured and unorganized items from huge collections to satisfy an information need. Museum information is complex due to the variety of materials that must be categorized, such as context text, photos, videos, and 3D digital objects. Blackaby and Sandore~\cite{blackaby1997building} create a bridge from guided web exhibits to unguided knowledge discovery through the construction of IR systems that hold heritage content. Koolen et al.~\cite{koolen2009information} show that the problem of disclosing cultural heritage information can be naturally presented as an IR problem. Motomura et al.~\cite{motomura2000generative} introduce an information retrieval system, ART MUSEUM, and show an interactive learning mechanism through their work. Wang et al. ~\cite{wang2007grid} propose a natural language-based IR system in the digital museum to popularize knowledge and encourage the excitement of discovery. Pechenizkiy and Calders ~\cite{pechenizkiy2007framework} focus on the problems of efficient learning visitor preferences of online and offline settings and present a framework for personalized access to the cultural heritage content in museums. Addis et al.~\cite{addis2003sculpteur} present the design and implementation of an innovative IR system that provides searching, navigating, and querying multimedia information held by museums and galleries. Although there are many related works discussing the adaption and innovation of IR systems in museums as the above ones, almost all of them are within the realm of formal learning and experiencing museums. Few works address systems for pre-learning, which are becoming significant with the current explosive growth of information~\cite{ally2004foundations}.

\subsection{VR Systems with Information Retrieval}
Immersive, interactive, and imaginative are three attributes of VR technology~\cite{radianti2020systematic}. It offers new multidimensional emotions and methods of human-computer connection~\cite{pan2006virtual}. It encompasses diverse capabilities such as scene construction, information resource construction, and user tasks. Museums have traditionally served the purpose of communicating knowledge~\cite{roberts2014knowledge}. Due to the recent growth of cultural tourism, internet platforms, and self-media user-generated content, the Internet is replete with information about museums and exhibits. A certain degree of pre-learning before the on-site experience will benefit a coherent understanding~\cite{liu2016identifying} of the corresponding society based on the artifacts and annotations shown in museums. It will likely bring added value for users to perform pre-requisite information retrieval~\cite{dumais2003stuff} in VR. Therefore, we first examine the VR-related works on information retrieval. Most of the research works are on technology frameworks such as user interface~\cite{brown1999developing} or eye tracking algorithm~\cite{mcnamara2018using}. There are only a few discussions in the fields of library and archiving applications~\cite{li2018research,hu2019application,fang2020visualization}. Although museums around the world have incorporated VR for years, research almost always focuses on adapting VR technology to promote engaging and meaningful experiences~\cite{wojciechowski2004building, li2016virtual,schofield2018viking,plecher2019mixed,puig2020lessons,harrington2020augmented}. To the best of our knowledge, there are no works on VR-based information retrieval for museums.    

\subsection{VR Visualization Branching Systems}

We executed a series of keyword searches in the primary computer science databases such as ACM and IEEE digital libraries and through Google Scholar for a broader subject matter of our design system. From this search, only a tiny part of MacLennan's work~\cite{maclennan2007design} relates to our topic. But some industry projects serve as good support; For instance~\footnote{Immersed: https://www.immersed.com/} offer a private Workspace and spawn up to five virtual monitors from the computer into VR without any additional hardware. MacLennan points out that humans have to handle ever-increasing amounts of information and that anything that can help us "get a handle" on it could be invaluable. When immersive technologies have been used in an information retrieval context, there is little evidence that their design has been influenced by user studies or knowledge of user preferences. MacLennan's work thus intends to address this apparent gap in the research. The study examines the question of user preferences for a virtual world designed to facilitate access to information. Alan proposed the question based on extensive research but did not specify any detailed solution. Our study moves one further step by examining user preferences in VR through the system designed to facilitate access to and organize information for pre-learning.

\section{Scenario}
\label{sec:scenario}

Our proposed pre-learning system is specifically designed to assist users in retrieving and acquiring information before visiting a museum. In this scenario, users are faced with a vast amount of tour-related information from various sources, such as official websites, social media platforms like Twitter, and recommendations from other visitors. Each user requires access to specific services and unique narratives, leading to the need for personalized recommendations and customization. However, traditional 2D interfaces have limitations in terms of screen size and information platform interoperability, making information retrieval for a tailored user experience challenging. 

To address this challenge, our system leverages virtual environments to present information interfaces as multiple screens. With the ability to browse multiple screens simultaneously and interact with them, users can efficiently navigate between websites and access customized and relevant information. This parallel browsing approach overcomes the need for users to navigate between multiple websites or manage excessive windows on the internet.

In our study, we conducted experiments using a large local museum as the test subject. Participants utilized Oculus VR headsets to interact with the system. The VR system loaded pre-learning information and presented it to users. Using Oculus handles, users could perform various actions such as scrolling, clicking, zooming in and out. The interaction allowed for content prioritization by dragging and placing them in different positions, similar to visual mapping tools like Xmind\footnote{\url{https://xmind.app/}} or Miro. Through these activities, users could collaboratively view content across WebViews in a highly customized and flexible manner.

\section{Formative Study}
\label{sec:formative}

Prior work on information retrieval systems for museum touring is lacking. To guide our design, we conducted a formative study to understand the challenges users face when using traditional web browsing on desktop computers and in VR.

During the formative study, participants were asked to use the VR default browser and a 2D screen (specifically, Safari on a laptop) to plan a visit to a large local museum within a time frame of 15-20 minutes. They were instructed to gather information about the museum, and exhibitions, and create a custom tour schedule using both settings. Participants were informed that they needed to find information such as routes, tour itineraries, tickets, opening hours, exhibition themes, interesting exhibits, and artists. The experiment would end if participants either found all the required information or failed to do so within 20 minutes. Participants were asked to jot down the relevant information they found on an A4 tissue paper for later reference. The task was customized based on each user's browsing habits, and the primary browser displayed relevant information pages based on the user's selections and clicks.

The experiments revealed that while users were proficient in using 2D browsers, it was challenging for them to simultaneously retain and browse multiple pages with relevant information. VR browsers had additional drawbacks, such as remote control operation and interaction, page zooming, and page prioritization. Based on our formative research findings, we established the motivation and design guidelines for developing the collaborative visualization VR PreM+ system.

\subsection{Participants and Apparatus}
We performed a between-subject evaluation with two conditions.
We recruited 24 participants (12 participants per condition), whose ages ranged from 21 to 59 years (M = 28.42, SD = 10.06) with 25\% identified as female. All participants were holding a Bachelor's degree or higher, and their mean familiarity with VR was 3.5, SD = 1.8 (measured by a 7-Point Likert scale). 

Participants were recruited from a local university. They all had normal or corrected-to-normal (using contact lenses or wearing glasses) normal visual acuity. None had experience with the PreM+ system used in the experiment or a similar system. The experiments were conducted in a confined university laboratory without other obvious distractions such as lights, noise, etc. No participants reported elevated susceptibility for motion sickness when queried using the Motion Sickness Susceptibility Questionnaire Short-form (MMSQ-Short) \cite{golding2006predicting}.

\subsection{Formative Study: Method}

The study, under operator supervision, began with a training session to familiarize participants with the system. They were informed of their right to withdraw at any time. The training involved the practical application of the system with an emphasis on addressing challenges faced by less experienced participants.

Participants were assigned to either the 2D desktop or VR browser group and tasked to prepare a museum tour following a specific scenario (Section~\ref{sec:scenario}). The task required them to retrieve key information within 15 minutes. We supplied A4 paper for note-taking, with the notes collected for subsequent analysis. We informed them of a subsequent quiz on the retrieved information.

User performance and preference were measured via time spent on task completion and a questionnaire that included NASA-TLX \cite{harris1993effect} and User Experience Questionnaire (short version) \cite{schrepp2017design}.

Post-study, we conducted semi-structured interviews~\cite{kallio2016systematic}, recording feedback about the experience and the use of 2D and VR for museum touring. Transcripts of these interviews were analyzed in Nvivo~\footnote{\url{https://www.qsrinternational.com/nvivo-qualitative-data-analysis-software/home}}. Based on this analysis, we made system adjustments. Other collected data included biosignal detection, interview photos, participants' notes, and task completion records.

\subsection{Formative Study: Findings}

The findings of our formative study are divided into the time spent in two situations, the results of the NASA-TLX, UEQ-Short questionnaire and the results of semi-structured interviews conducted with participants.

\subsubsection{User Performance}

Participants using 2D screen-based ($Mdn=808.67s$) were faster than those using VR headset default browser ($Mdn=939.67s$), confirming an already prevalent hypothesis that 2D-inspired interaction methods do not work well in immersive environments~\cite{lam2019m2a}.

\subsubsection{User Preference}

\paragraph{NASA-TLX}
We employ an independent-samples t-test with two conditions as the 2-level within-subjects variable. There is no significant effect of the browsing technique on the overall workload for the 2D ($M=21.58$, $SD = 3.80$) and VR ($M=21.08$, $SD = 3.75$) conditions ($t(22)=0.32$, $p=.749$). 
For the NASA-TLX subscales,  VR barely escapes statistical significance in terms of Effort compared to 2D ($t(22)=1.82$, $p=.083$). There is no significant effect of the browsing technique on the Mental ($p=.543$), Physical ($p=.583$), Temporal ($p=.292$), Performance ($p=.388$), Frustration ($p=.478$) subscales.

\paragraph{UEQ-Short}

The analysis yields no significant influence of the browsing technique on any of the UEQ-Short subscale: \textit{pragmatic quality} ($p=.870$), \textit{hedonic quality} ($p=.497$). VR ($M = 36.83$, $SD = 7.11$) had a higher UEQ-Short score than 2D ($M = 35.42$, $SD = 5.35$).

\subsection{Observations}

Besides administering the questionnaires, the operator also observed the participants' behavior when using the desktop 2D browser or the VR browser. When using the 2D browser, participants often closed pages they did not think were important at first and had difficulty retrieving them later. When using the VR headset default browser, operational errors often occur due to imprecision in handling the VR joystick.

\subsubsection{Semi-Structured Interview}
Following the formative study, we asked participants to share their experiences with 2D or VR and offer some suggestions. All participants in the 2D Group (P1-P6),  think that the existing 2D web pages can let them find basic information, but most of them (P1, P2, P5, P6) think that traditional 2D search is not very convenient. P3, P4 and P5 have the habit to use more than two screens in daily life. They think the dual screen can improve efficiency in daily work. User 1, an Artist, says he \textit{``usually follows a lot of tourism websites and social media''}. User 4 want a special platform that combines basic official information but allows other users to share experiences and tips. \textit{``It (multi channel search) is necessary''} (P5). All of them felt that if there were more screens and branching systems, the search efficiency would be better. P3 thought it would be very helpful to have multiple screens in VR and be able to put the most focused information in the middle.

Most participants in the VR Group (P7, P8, P10, P11, P12) think that using the default browser in a VR environment is not as convenient as using a traditional computer to browsing information. The main reasons  are: 1. Using handle controller was too slow to enter information on the virtual keyboard (P7, P8, P10, P12). 2. VR devices are uncomfortable to wear, especially for people who wear glasses (P11, P12). 3.  It was not easy to record useful information in VR (P12).  Most of them (P9, P10, P11, P12) think that multi-screen in VR would be more comfortable and convenient. \textit{``Multi-screen is definitely more comfortable to use than one screen.''} (P9).

\section{VR PreM+: An Immersive Pre-Learning Branching Visualization System}
\label{sec:design}

Our design has three main goals: (1) to provide users with immersive, situational experiences, (2) to enable advanced information retrieval and personalized route planning in a virtual environment, and (3) to facilitate the efficient acquisition of necessary information~\cite{sutcliffe2019reflecting}. To achieve these goals, we propose and implement the PreM+, which utilizes multiple 3D web windows and related 3D objects in a virtual scene to improve the user experience compared to traditional PC browsers and the default Oculus Quest 2 browser.

\subsection{Design Requirements and Decisions}
We developed a VR system with hierarchical information classification to achieve the outlined design goals. It facilitates navigation across various channels and platforms, employing multiple-screen displays. Key features are:

\begin{enumerate}
\setlength\itemsep{-0.1em}
\item \textbf{Multiple-screen display support}: Unlike traditional displays, screens in the VR environment are customizable in size, number, and layout, enhancing user immersion and efficiency by allowing parallel operations on different interfaces.

\item \textbf{Multi-platform navigation and organized information}: This feature addresses users' challenges in quickly retrieving accurate information.

\item \textbf{User-customizable information organization}: A branching system enables targeted visualization of relevant information. Users can customize content, order, and placement of pages, promoting collaboration and information exchange across various devices.

\item \textbf{Cross-platform applicability}: Users can collaborate through multiple mediums without a VR platform. The system supports cross-device collaboration, thus enabling information sharing through different platforms and devices. 
\end{enumerate}

\subsection{User Scenario and Interaction Design}
Our VR PreM+ system, developed using the Unity engine\footnote{\url{https://unity.com/}} and Vuplex 3D WebView plugin\footnote{\url{https://github.com/vuplex/oculus-webview-example}}, offers a simulated environment for pre-learning. This system is utilized to present a virtual exhibition hall, aiding users in assimilating information prior to a real-world museum visit, as depicted in Fig. \ref{fig:teaser}. We utilized a modern art museum scene\footnote{\url{https://assetstore.unity.com/packages/3D/environments/art-gallery-vol-9-modern-pavilion-202478}} as our pre-learning system's backdrop.

\begin{figure}[t]
    \centering
    \includegraphics[width=1\linewidth]{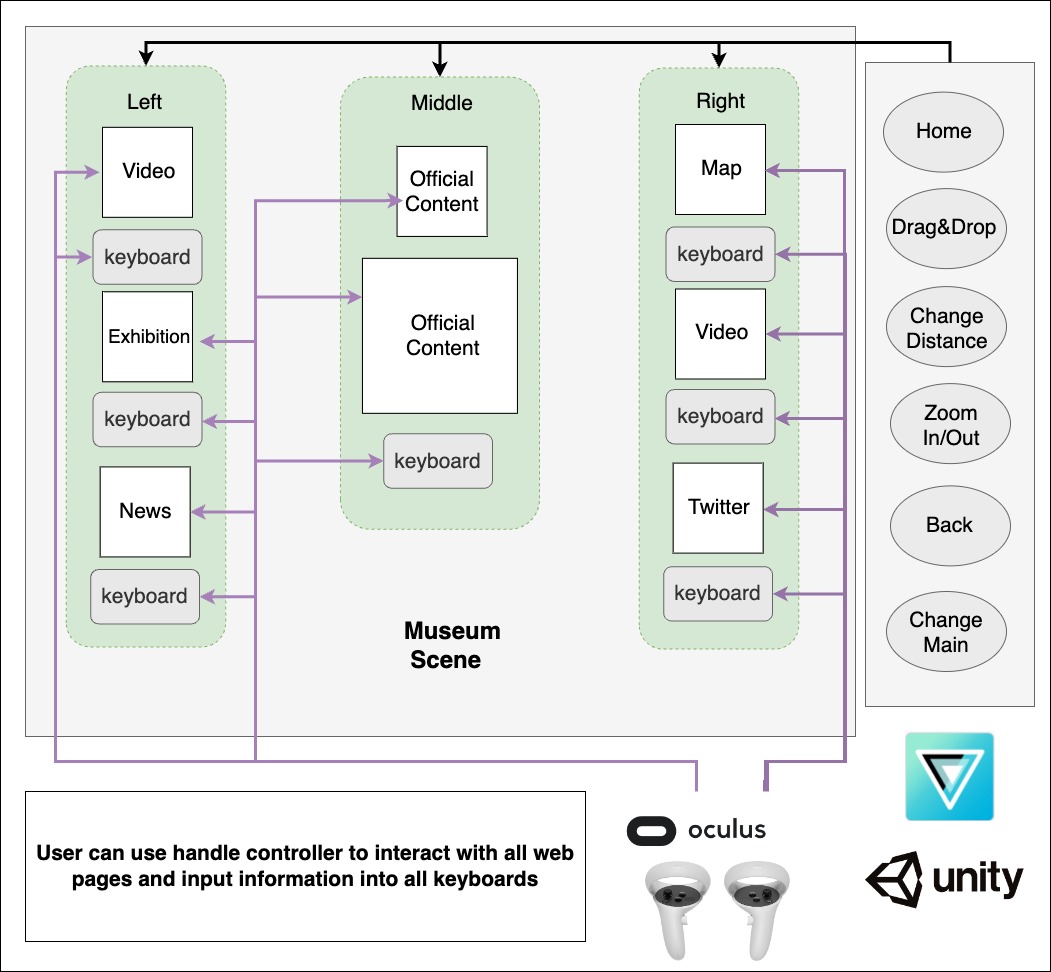}
    \caption{System Architecture. The designed system relies on the 3D WebView of Vuplex and Unity. The position of pages and the content of each page can be customized, and this diagram is just an example.}
    \label{fig:sysarch}
\end{figure}

The VR PreM+ system facilitates three levels of information classification. The first level gleans information from a notable local museum's official website\footnote{\url{https://www.mplus.org.hk/en/}}. At the second level, users can tailor the display of web content supplied by various platforms. In the third level, users can further customize cross-platform navigation, presenting useful tips for their visits.

We have designed the system with a multiple 3D WebView page layout. The default interface offers a three-column arrangement: left, middle, and right. The system can also accommodate any number of columns, but our experience with the Oculus Quest 2 shows that the resolution limits readability to three columns. The columns' content guarantees users can compare information with minimal effort.

Our system employs the classic VR interaction mode based on the Oculus Touch Controllers\footnote{\url{https://docs.unity3D.com/560/Documentation/Manual/OculusControllers.html}}, which is easy for users who are familiar with basic VR operations to get started. Both controllers are symmetrical, making our system user-friendly for both left-handed and right-handed individuals. We have equipped the system with features that permit users to adjust the distance between the right or left column of web pages and their eyes. Users can also freely drag and drop the selected webpage. Similarly, users can modify each page's resolution. This functionality, coupled with the ability to control the content displayed on the web pages, offers users a high degree of customization. A unique feature of the VR PreM+ system is the flexibility in controlling the amount of content displayed and its position. As depicted in Fig. \ref{fig:teaser}, users can exhibit up to 8 WebView pages at once, surpassing the capabilities of a laptop or the Oculus Quest 2's default browser.


\begin{figure}[ht]
  \centering
  \includegraphics[width=\linewidth]{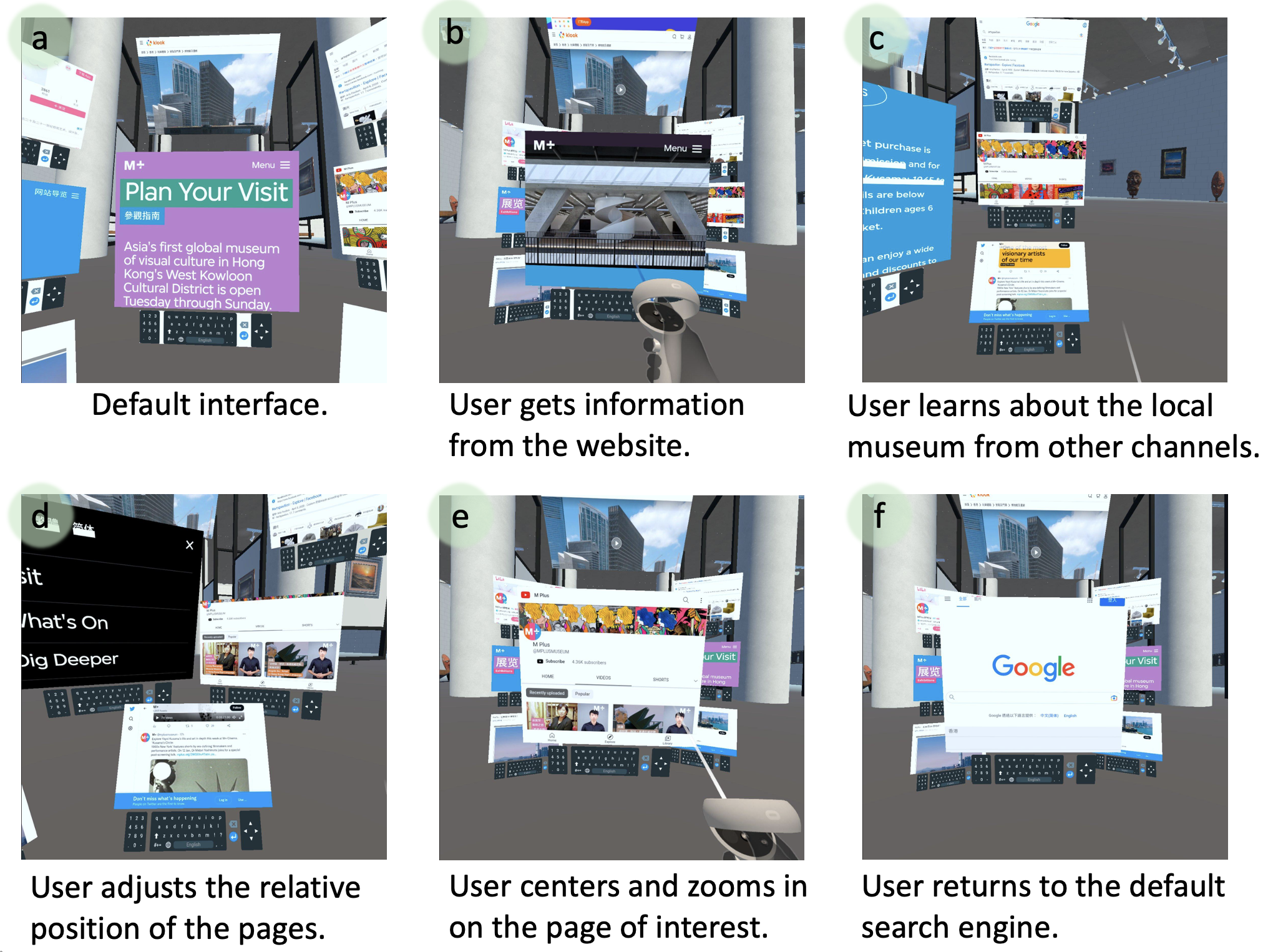}
  \caption{Storyboard of a possible scenario for information gathering in our system. a) to f) show the system interface in VR when the user operates in order. In our storyboard, we take the pages on the right side as an example. Performing a series of operations on the left side is exactly the same.}
  \label{fig:storyboard}
\end{figure}

In a typical user scenario, as shown in Fig. \ref{fig:storyboard}, users are presented with eight WebView pages, with the museum's official website and ticketing site displayed prominently. Users can then adjust the main screen's position and zoom level, and reorganize the screens around it. Searches can be conducted using the default search engine, with a 3D keyboard included. Users can drag the important pages to the main screen, prioritizing their viewing order.

\subsection{Implementation}

The VR PreM+ was developed in Unity3D and built to run in Oculus Quest 2, which contains two wireless handheld controllers and one head-mounted display. Considering the amount of information that ordinary users need to refer to simultaneously, we set the maximum number of pages per column to 3 by default. The main 3D WebView pages in the middle column are 80cm (width) $\times$ 60cm (height) in size. Other 3D WebView pages are initialized with a size of 60cm (width) $\times$ 40cm (height). All pages are initialized to a resolution of 1300. 
The relative initial position of all pages from the eyes is 0.8m. The range of adjustable distance is 0.4 meters to 2 meters.
To improve the user's visual effects, we use \textit{``Transform.LookAt()''} API in Unity\footnote{\url{https://docs.unity3D.com/ScriptReference/Transform.LookAt.html}} to rotate the out-of-center WebView page at an angle to align with the user's eyes. All 3D WebView pages are facing the user after being dragged and changed position, so users can experience our system in various positions, even if they are lying on their back on the couch. In the 3D space of VR, the rotation adjustment of WebView pages is very effective. The proper angle deflection can make users more comfortable browsing information and give them a wrap-around style.

\section{User Study}
\label{sec:evaluation}

In our usability testing study, we aimed to systematically evaluate the performance and user experience of our system, using a similar workflow and structure to that of our previous formative studies. Participants were first required to complete a demographic information questionnaire and undergo training on the use of our system. Prior to the formal user study, participants received pre-training on how to interact with the system's various Webviews. The user study consisted of tasks related to obtaining information about a local museum and creating an optimal visiting route. Instead of comparing scores across tasks in a ``fair" manner, we focused on comparing measures of underlying performance between the user study and the formative studies.

\subsection{Hypotheses}

In order to evaluate the performance and efficiency of PreM+, we made the following assumptions before the user evaluation.

\textbf{H1} Our system offers improved efficiency and comparable information and functionality compared to 2D screens or VR headsets with their browsers. This is achieved by allowing the user to remain highly focused on the assigned task within the virtual environment, rather than constantly switching between pages.

\textbf{H2} The PreM+ system facilitates the creation of an optimized and personalized tour path for the user. This significantly reduces the operational limitations associated with traditional single-screen browsing. Traditional single-screen-based browsing often leads to the retrieval of irrelevant information, making it challenging to develop an optimized path.

\subsection{Participants and Apparatus}

We recruited 18 participants from a local university, whose ages ranged from 21 to 36 years (M = 25.34, SD = 3.88) with 50\% identified as female and all had a Bachelor's degree or higher. The mean familiarity with VR devices of our participants was 4.5 (SD = 1.4, measured by a 7-Point Likert scale). All had normal or corrected-to-normal (using contact lenses or wearing glasses) visual acuity. The experiment was conducted in a university lab. Note that none of our participants had previous visiting experience at this Museum and no participants reported elevated susceptibility for motion sickness when queried using the Motion Sickness Susceptibility Questionnaire Short-form (MMSQ-Short) \cite{golding2006predicting}. To facilitate the subsequent analysis, we labeled these 18 participants as P25 to P42.

\subsection{Evaluation Metrics}

We recorded the time cost (task completion time) for each participant to quantitatively compare the learning and exploration time of participants using our system. We measured and collected data describing users' preferences for the methods, including subjective feedback (user experience, workload, immersion, and motion sickness). We also conducted semi-structured interviews at the end of the studies to receive more subjective feedback and comments, especially considering future improvements.

\begin{figure*}[ht]
  \centering
  \includegraphics[width=1\linewidth]{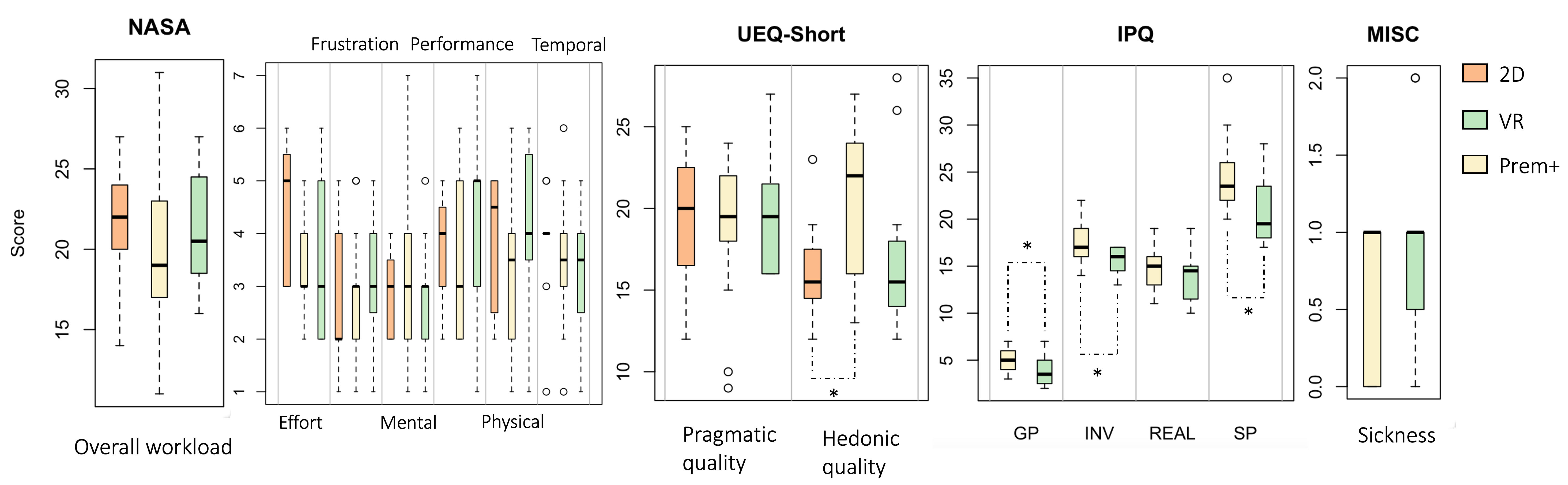}
  \caption{User performance. Data was obtained through four questionnaires: NASA, UEQ-S, IPQ, and MISC. Among them, IPQ and MISC were analyzed only in the VR environment. ($P<.05$ (*))}
  \label{fig:userstudy}
\end{figure*}

User Preference was measured by two questionnaires compiled from the NASA-TLX \cite{harris1993effect}, User Experience Questionnaire (short version) \cite{schrepp2017design}, Igroup Presence Questionnaire \cite{schubert2001igroup}, 11-point MIsery SCale (MISC) \cite{bos2015less}, and semi-structure interviews \cite{kallio2016systematic}.

\subsection{Study Results}

We compare the results of the user study with those of the Formative Study. The Brown-Forsythe test ($p=.69$) shows unequal variance across conditions due to inconsistent sample sizes. Therefore, we perform non-parametric analysis using the Kruskal-Wallis H test.

\subsubsection{User Performance}

\begin{figure}[ht]
  \centering
  \includegraphics[width=0.7\columnwidth]{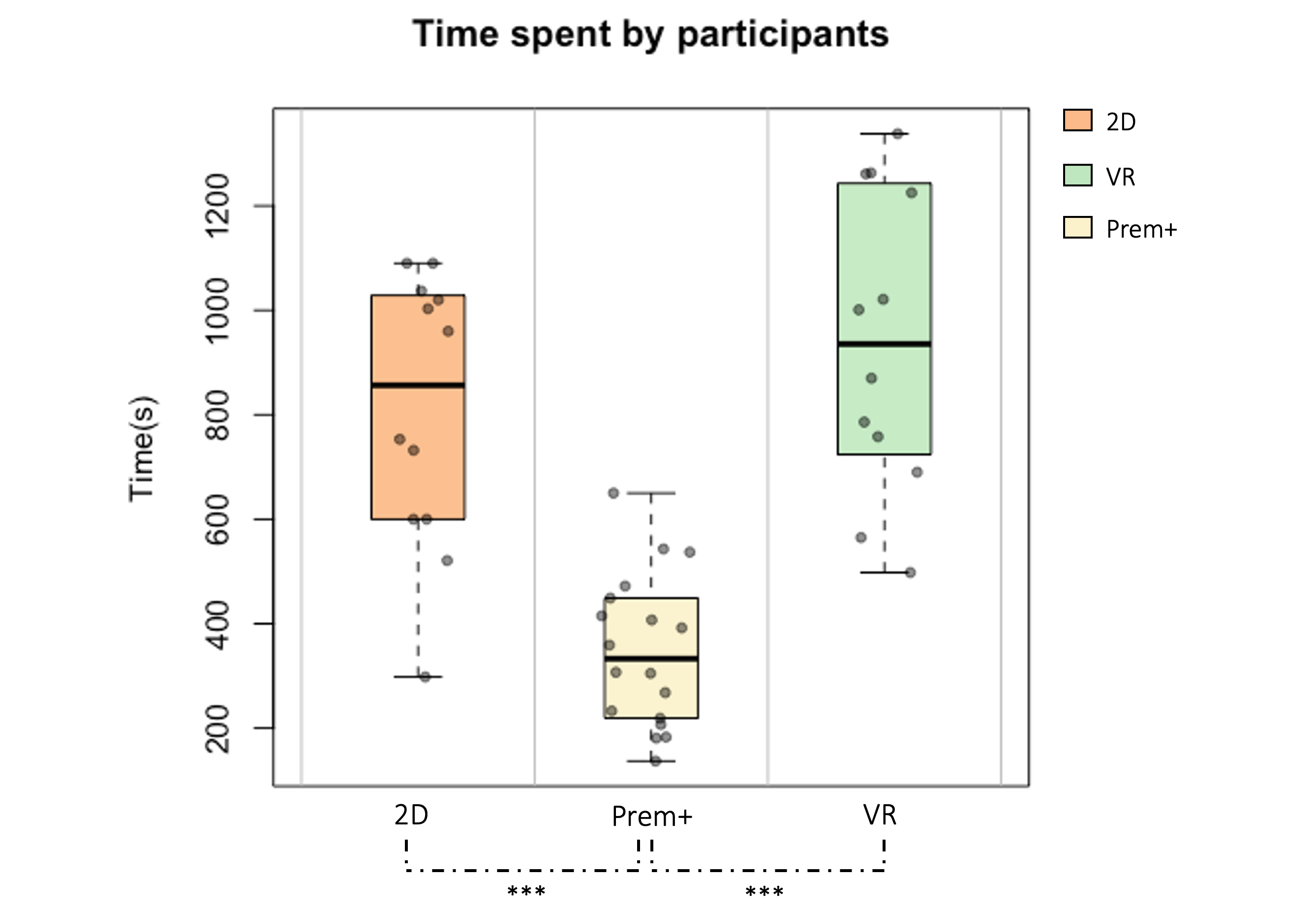}
  \caption{The time taken to complete the tasks by the participants in the three conditions. The figure shows that the time taken by participants using the PreM+ system was significantly less than 2D and VR. ($P<.05$(*), $P<.01$(**), $P<.001$(***))}
  \label{fig:userstudytime}
\end{figure}

Figure~\ref{fig:userstudytime} shows the time taken to complete the task between the 2D browser, the VR default browser, and VR PreM+. There is a significant difference between the different conditions  $\chi^2(2)=25.406, p<.001$, with a mean time of $808.67(s)$ for 2D, $939.67(s)$ for VR and $348.00(s)$ for PreM+.  Post hoc analysis shows that the time spent in PreM+ is significantly lower than in 2D and VR (both $p<0.001$). The results of the study validate \textbf{H1} that our system is more effective compared to a 2D screen and a VR headset with a built-in browser.

\subsubsection{User Preference}

We evaluate user preference with data obtained from four questionnaires, NASA, UEQ-S, IPQ and MISC. For details, see Fig.~\ref{fig:userstudy}.

\paragraph{NASA-TLX workload}
The overall workload required for three conditions, PreM+ ($M=19.89$, $SD=4.89$) is slightly less than 2D ($M=21.58$, SD=$3.80$) and VR ($M=21.08$, $SD=3.75$). However, there is no statistically significant effect of the technique on the overall workload ($\chi^2(2)=1.82, p=.403$). We also do not find any significant effect of Mental ($p=.809$), Physical ($p=.354$), Temporal ($p=.401$), Performance ($p=.440$), Effort ($p=.067$), Frustration ($p=.578$) subscales.

\paragraph{User Experience Questionnaire - Short (UEQ-S)}
The conditions significantly affect the \textit{hedonic quality} subscale of the UEQ-S ($\chi^2(2)=6.608, p=.038$), with no significant effect on the \textit{pragmatic quality} subscale ($\chi^2(2)=.043, p=.979$) with 2D ($M=19.42$, SD=$3.82$), VR ($M=19.67$, SD=$3.55$) and PreM+ ($M=18.89$, SD=$4.11$). Post-hoc tests confirm that PreM+ ($M=20.45$, SD=$4.41$) cause a significantly ($p<.05$) higher hedonic quality than 2D ($M=16.00$, SD=$3.05$), but not significant compared to VR ($M=17.17$, SD=$4.99$).

\paragraph{Igroup Presence Questionnaire (IPQ)}
A Mann-Whitney U test shows significant differences in Considering general presence (GP), Spatial presence (SP), and Involvement (INV) subscales for the PreM+ and VR conditions. 
PreM+ scores significantly higher than VR on all three subscales of GP with a mean score of 5.06 (STD=1.61) vs 3.83 (STD=1.64) for VR  ($U = 158$, $P = .030$), SP with a mean score of 24.56 (STD=3.85) vs 21.00 (STD=3.69) for VR  ($U = 166.5$, $P = .012$) and INV with a mean score of 17.39 (STD=2.23) vs 15.58 (STD=1.56) for VR ($U = 153.5$, $P = .049$). There is no significant difference between VR ($M=13.83$, $SD=2.66$) and PreM+ ($M=14.78$, $SD=2.13$) in terms of the Realism (REAL) subscale ($U = 132$, $P = .315$).

\paragraph{Motion Sickness MIsery SCale (MISC).}
The results of the Mann-Whitney U test show that there is no significant difference between PreM+ ($M=.56$, $SD=.51$) and VR ($M=.92$, $SD=.67$) on Motion Sickness ($U = 77$, $P = .144$).

\subsection{Interviews}

In this section, we conduct a thematic analysis of 12 participants' semi-structured interviews through Nvivo\footnote{\url{https://www.qsrinternational.com/nvivo-qualitative-data-analysis-software/home}}. We focus our analysis on four themes: scenario design; insights; suggestions; and additional comments from different stakeholders.

\subsubsection{Semi-Structured Interview Format}

Through this study, we expect to better identify the strengths and weaknesses of VR branching system design and understand the extent to which information retrieval in a VR environment enhances the user experience ~\cite{rubino2015integrating}. 
We introduced our project to a panel of 12 participants, distinct from the experimental panel in the formative (Sec.~\ref{sec:formative}) and final study (Sec.~\ref{sec:evaluation}),  composed of one architect  (P19), one Engineer (P21), and ten members of a general audience(student) who have a general interest in museum visits (P13, P14, P15, ...and P24).

The semi-structured interviews focused on the participants' overall feelings about the system, their feedback on their use of the VR interaction system, how they envisioned the scenario that best suited them, and how they understood the tasks that needed to be accomplished and reflected on how their scenarios could have better designed the system. We found that participants shared their experimental experiences in two ways: their intuitive feelings and the visual effects and interactions that they expected.


\subsubsection{Scenario Design}
Visitors appreciated the experience of being in the virtual museum environment through a head-mounted device for museum information retrieval. The amount of preview information contained on multiple screens is far more significant than the amount of information contained on one screen. It offers a primary and secondary information hierarchy and a comparative analysis of different information platforms. Such an operation helps them to understand various information about the exhibition they are about to visit in the shortest possible time, such as the exhibition theme and artists, and to understand the artist's narrative through this process. For example, P33 commented, \textit{``When I went to the Hong Kong Palace exhibition with a friend in July this year, my lack of understanding of the interior structure of the pavilion resulted in missing many exhibits and individual sections.''} She also said that it was difficult for regular art exhibition-goers like herself to miss any of the exhibits completely. P34 also said that she often did not know the theme of an exhibition until after she had seen it. Information retrieval through a pre-learning branching system like PreM+ can form a complete viewing plan and give a vivid image of the exhibits that are about to be physically seen. Through this process, viewers form a storytelling picture that improves their understanding of the exhibition and their impressions of the exhibits. (P25, P26, P27) stated that they have also visited some exhibitions with VR devices. Using VR devices to preview or understand the information about viewing the exhibition in a real museum space gave them a meaningful and enjoyable interactive experience that filled the gaps that 2D screens could not. The information comparison and prioritization features were essential, and the physical museum scenes gave them a more immersive feel. However, they point out that the application's main advantage is information retrieval, matching, and sorting before viewing. Scene and immersion may not be necessary, but (P29, P31, P38) also indicate that if background music related to the exhibition theme is added, it can remove the distractions caused by the surroundings, allowing users to focus more and reach a better retrieval efficiency.

\subsubsection{Insights}
Most users in the semi-structured interviews felt that using the PreM+ branching system for the tour was more personal and relaxing than the traditional application (2D). A few participants (P33) found it challenging to understand how to use the visual multi-screen branching system in a virtual environment because they were unfamiliar with the technology. However, they quickly understood this after we explained our design process with examples. Many of them (P25, P26, P29, P32, P34) were impressed with the interactive function (zoom in, zoom out, pull in, pull away) as it improves convenience and highlights the priority of information.

(P25, P26) liked the following function of the PreM+ branching system because the multi-screen still follows as the head swings. It helps the viewer get on the information retrieval track quickly and correctly and helps the user find the main functional areas. (P30, P32) say that very cool visual effects should be convenient for young users. For (P31), if the system focuses on effective information gathering before a trip to the museum through such efficient information retrieval to cope with it, it should concentrate on the key information they want to emphasize and the information that is difficult to convey (e.g., important exhibits and exhibition opening times, whether they need to buy tickets or not). Therefore, applying the proposed system in the museum can be seen as a communication design method. On the other hand, (P25, P26, P27, P29, P36, and P37) think that exposing too much information without the relevant understanding may prevent them from exploring further. This difference between people from art backgrounds and general audiences needs to be further explored to create a better experience of information retrieval within the virtual space and to guide users through PreM+ so that they can better understand the exhibition before visiting it and develop their path to visit.

\subsubsection{Suggestions} 
We received valuable suggestions about the operation of the system. First, several audience members (P25, P26, P27, P29) mentioned that the function keys on the controller were significantly more precise than dragging operations but also lacked an innovative interactive experience. P25 appreciated that the PreM+ system was operated within the virtual space of the museum for a better immersive experience. However, they noted that the experience would have been more interesting if it had been related to the exhibition to visit. On the other hand, some participants, such as P33, suggested that the virtual environment of the museum as a backdrop can be overwhelming when using the PreM+ system by adding unnecessary information. They suggested switching to a more open view. P35 and P25 suggested that a clear and explicit setting process and workflow must be improved for personalized exhibition path setting. P31 mentioned that the controller is too sensitive, and sometimes a gentle dragging will produce a page slide so fast that they miss important information. Finally, P31 and P33 had the following suggestions for further user needs. The system can be applied to the other exhibition previews, where scenarios can further explain the artist's creative concept and production process, creating more intuitive interactions~\cite{livrcaptcha,li2023swarm} with the participants in the previews. They also suggested creating VIP personalized and customized tour paths through PreM+ before visiting an exhibition for a more personalized route and experience. Support for personalization will be a significant future trend for museum previews, where visitors can customize scenarios for a more professional and specific tour. Other participants, such as P28, stated that the experience of VR systems could be multi-sensory and multi-modal rather than limited to visuals. P34 stated that if exhibition paths are customized through PreM+, later after visiting the exhibition, one can also review the paths or exhibits they have visited through VR, and even design interactive gaming experiences may be an enhanced experience before the age of the metaverse \cite{xu2020virusboxing,li2021myopic}.

\subsubsection{Additional Comments from Different Stakeholders}
The architects (P31) and general participants (P36) felt that this preview information helped them better understand the exhibition's information. No important exhibits or content were missed. However, they would like more innovative or abstract interactive designs to absorb more viewers in the virtual environment of VR. The audience (P29) was impressed with this new approach to previewing. He also mentioned that this work could be a good tool for commercial museum research and advertising purposes, especially for improving awareness of exhibition information and understanding of exhibition content. In addition, P34 said that PreM+ might be set to different levels to accommodate differences in the artistic background of the audience. For example, more information about the artist's background and important works may be provided to the general public. For art-trained viewers, the tour should provide different types of in-depth information retrieval, e.g., automatic scheduling based on the genre and style of the exhibit. Their differences give the user more freedom to develop their own story while accessing basic information about the exhibition. Finally, (P33, P36) complained that although they felt like this emerging approach was visually and experientially eye-opening. However, the multiple screens in VR sometimes made them feel overwhelmed. Of course, this is also related to their habits of using computer screens, cell phones, and other single-screen devices. We also recommend choosing the best number of screens for better experimental results after a later review of the number of screens. This comment shows that personal habits also need further consideration.

\section{Design Strategies}
\label{sec:strategies}

This section summarizes strategies to guide the design of future pre-learning systems, especially in an immersive environment. Our strategies are exploratory in the sense that has been demonstrated primarily through the PreM+ system and user studies we have designed. As such, we have not yet evaluated them through follow-up research, but we plan to hold design workshops and in-depth collaborations with several local museums to gather further empirical evidence that these strategies are helpful to designers \cite{patibanda2023a,patibanda2023b,floyd2021limited,mueller2023}.

\subsection{Display}

\subsubsection{Display as a Branching System}
This strategy concerns the extent to which the system supports visitors, or users, by presenting them with multiple different sources of information simultaneously in the form of a branching system. It draws on the study of visualization, providing separate interactable windows for many different kinds of information to be collected and organized. In our user study, almost all participants noted that displaying information in a branching stream was ``very useful'' for them to find different resources at the same time quickly. This strategy can even be widely applied, even in non-VR 3D environments. It can also extend to 2D environments, in the case of large displays with very high resolution. However, the VR environment allows exploiting a third dimension to better sort the content, for instance by spreading the grid layout over 360 degrees around the user.

\subsubsection{Display as an Immersive Interactive System}
This strategy focuses on the benefits that the immersive environment we provide in PreM+ brings to the user's pre-learning. The virtual environment provided by VR devices is much more immersive than the interaction brought by a 2D screen. The user is fully engaged in using the device to exclude redundant distractions, such as interruptions from other programs. Our formative study confirms this finding. Most participants noted that this VR interaction (i.e., VR default browser) felt novel to them and was not susceptible to outside interference, which was very useful for gathering information.

\subsection{Interaction}

Interaction is crucial in VR systems because it bridges the gap between behaviour in VR and real-world interactions \cite{barathi2018interactive}. However, there are challenges in using handles for interface interaction in VR, particularly for information retrieval tasks that require point-and-click \cite{otto2019virtual}. Our study found that displaying too many screens can overwhelm users \cite{vatavu2014visual,vatavu2015evaluating} and distract them during information retrieval \cite{ cruz1993surround}. Therefore, designers must find ways to balance the number of screens and user attention. It is also challenging to effectively support users when collaborating between web pages. To address this, we focused on providing customers with more customization options, and the ability to move freely within each page and modify the page's content.

\subsection{Humanized Design}

Our design focuses on enhancing information-gathering effectiveness and user perception of the system. We aim to introduce more humanized elements to improve the system's usability and user-centered experience. Based on experimental data and user feedback, we plan to iterate and enhance the system in the following areas, with the goal of creating a more user-friendly and user-centric experience. We propose four design considerations for the VR system:

\subsubsection{Matching Target Information Search and Returned Content}
Users can browse websites in a virtual space, and when they search for keywords or phrases, relevant content will be displayed through a new AR 2D interface or 3D graphics \cite{lam2021a2w}. For example, if a user searches for "3D modeling software," the system would display a 2D augmented Wikipedia website related to "3D modeling software," open different 3D modeling software programs, and present "3D modeling software" icons in a hybrid interface to support effective learning.

\subsubsection{Identifying Content Formats for Intuitive Browsing}
The system supports classifying and filtering content to present it in various formats, such as 2D interfaces and 3D images. The interface can display content information in a combination of different media formats, including text, images, audio, and video, based on conventional HTML5 design. The selection of formats is determined by user interaction, user information (e.g., user history, preferences, and temporal state), and the nature of the content.

\subsubsection{Personalized Composition and Display of Content}
To enhance the user experience and ensure a seamless workflow, the system compiles and presents content in a personalized manner. This step involves filtering and selecting XR content based on user information mentioned earlier. The XR environment interacts with the user in real-time, delivering relevant XR content whenever and wherever needed. Intelligent algorithms are employed to understand the user's preferences and personalize the content presentation.

\subsubsection{Remote Collaboration}
Users can engage in remote collaborative environments, allowing them to comment on the viewed information and assess its reference value with other users. This approach introduces a human virtual message layer, where each specific message is redefined by the user, adding a deeper level of interaction and feedback to the initial message.


\section{Limitations}
\label{sec:limitations}

In daily life and work, it is common to browse multiple pieces of information simultaneously. For example, doctoral students may use multiple monitors to read multiple papers, while others use multiple monitors to monitor stock market situations. Although our study focuses on museum tour planning, our VR PreM+ system can serve as a versatile pre-learning tool for various topics. It combines traditional information retrieval in virtual spaces, interactive multi-screen retrieval methods, and alternate virtual environment scenarios. As pre-learning is an informal learning process, it requires an attractive and entertaining design to enhance user engagement \cite{olsen2008interactive}. Some operations in our PreM+ system, such as comparing information by turning in the virtual space, differ from traditional 2D web page retrieval. Users need time to adapt to this new model through repeated use. We recognize the need to further refine and redesign these operations to improve the effectiveness of information retrieval and matching through VR.

During the formative study, participants using a 2D computer screen often preferred to use a mobile screen (e.g., iPad) for pre-learning rather than a computer screen. Mobile applications serve as containers for specific information, such as social media content on platforms like Facebook or Twitter, and align with their usage habits. In the semi-structured interviews conducted during the user study, many participants mentioned that the system improved the efficiency of information retrieval. However, due to the headset, note-taking was challenging. Participants suggested incorporating a note-taking function in VR, similar to the sticky note function found on Apple computers. However, convenient note-taking in immersive environments remains a significant challenge \cite{lee2020seen}, requiring dedicated studies for effective solutions.

Taking participant feedback into account, we plan to enhance the engagement techniques of the PreM+ system, including interaction methods with virtual environment objects. We also aim to improve existing interaction mechanisms within PreM+ to increase user engagement, knowledge acquisition, and categorization by prioritizing user-friendly design.

\section{Conclusion and Future Work}

In this paper, we explored the feasibility of information retrieval in VR by designing and evaluating VR PreM+, an immersive pre-learning branching visualization system. With the overarching scenario of planning a museum visit, we first conducted a formative study to evaluate the shortcomings of current default VR web browsers. The study shows that the 2D screen-based information retrieval capability is more effective than the VR browser, and users face many interaction difficulties in VR. Given the results of this study, we design and implement VR PreM+ as a virtual learning room where users can explore, extract, and classify pieces of information using the near-unlimited 3D space in VR. Our final user evaluation shows that VR PreM+ halves the time to gather information compared to both 2D and VR browsers. Users appreciated the functions to classify and arrange information in 3D. 



In the next research, we plan to conduct a co-design process to improve the system's operation and study the comprehensive design of hybrid interfaces further, considering recommender systems, human-computer interaction, and multimedia systems. 
Our future research will focus on creating a metaverse classroom that blends digital with physical. Concentrate on teamwork, co-creation, synchronous and asynchronous communication, and real-time communication. 
We hope that our prototype of the Hyper-Learning system and interfaces can encourage more artists and researchers who want to connect virtual learning content with physical spaces to further explore the usage of XR and metaverse as a tool for education and co-creativity.

\begin{acks}

This research was partially supported by the Extended Reality and Immersive Media (XRIM) Lab at HKUST and the National Social Science Foundation of China (20CJY045).

\end{acks}

\bibliographystyle{ACM-Reference-Format}
\bibliography{sample-base}

\end{document}